\documentstyle[twocolumn,prl,aps]{revtex}
\begin{document} 
\title{In-plane
Magnetic Field Induced Kosterlitz-Thouless Type
Metal Insulator Transition in Coupled Double Quantum Wells
}
\author{D. Z. Liu$^{*}$}
\address{
James Franck Institute, University of Chicago, Chicago, IL 60637
}
\author{X. C. Xie}
\address{
Department of Physics, Oklahoma State University,
Stillwater, OK 74078.
}

\address{\rm (Submitted to Physical Review Letters on 19 September 1996)}
\address{\mbox{ }}
\address{\parbox{14cm}{\rm \mbox{ }\mbox{ }
We study the localization properties in coupled double quantum wells with
an in-plane magnetic field. The localization length is directly 
calculated using a transfer matrix technique and finite size scaling analysis. 
We show that the system maps into a 2D XY model and undergoes a disorder
driven Kosterlitz-Thouless type metal-insulator transition depending
on  the coupling strength between the two-dimensional layers and the magnitude
of the in-plane magnetic field. 
For a system with fixed disorder,
the metallic regime appears to be a window in the magnetic field -- coupling
strength plane. 
Experimental implications of the transition will be discussed.
}}
\address{\mbox{ }}
\address{\parbox{14cm}{\rm PACS numbers: 71.30.+h, 73.20.Jc, 73.61.-r}}
\maketitle


\vspace{-0.5cm}

There have been long lasting interests to understand localization problem in
two-dimensional(2D) systems\cite{lee}.
According to the scaling theory of localization\cite{anderson:scal},
all states in a two-dimensional system are localized if only scalar random
potential is present. However, in the presence of a strong perpendicular
magnetic field, extended states appear in the center of disorder
broadened Landau bands and give arise to the quantum Hall effect\cite{prange}.
There has been report on metal-insulator transition (MIT) in the presence
of spin-orbit coupling in two-dimensional systems\cite{spin}. It is important
to understand the criteria for MIT in disordered 2D systems and corresponding 
universality classes for these phase transitions.  

From symmetry consideration, 2D localization problem can be classified into 
following universality classes: (1) orthogonal ensemble, where only scalar
random potential is present; (2) symplectic ensemble, such as the case
with strong spin-orbit scattering; (3) unitary ensemble, such as the quantum
Hall system, where the strong perpendicular magnetic field breaks the 
time-reversal symmetry. One interesting system belongs to the 
unitary class is the system with random magnetic field\cite{zhang}.
This system has attracted much attention recently because it is relevant to 
the studies of fractional quantum Hall systems at even denominator filling 
factors and  high-$T_c$ models with gauge field fluctuations.
In this problem, the time-reversal symmetry is {\it locally}
broken by the random magnetic field, whether this
symmetry-breaking is sufficient to induce metal-insulator
transition remains unclear despite extensive recent studies\cite{theory}.
However, Zhang and Arovas\cite{za} developed an effective field theory for
the random field problem, where they found the MIT in 2D random field is
a Kosterlitz-Thouless type transition\cite{moon}. 

In this letter, we study a different system where the time-reversal symmetry 
is also broken: coupled double quantum wells with in-plane magnetic field.
This system is
experimentally more accessible comparing with the
random field system and theoretical predictions
can be tested. We will demonstrate that the system maps into a 2D random field
problem and a Kosterlitz-Thouless (KT) type
metal-insulator transition arises depending on disorder, 
the coupling strength between the two-dimensional layers and the magnitude
of the in-plane magnetic field. With fixed
disorder, the metallic regime will be a window in the magnetic field -- 
coupling strength plane.
Experimental implications of the transition will be discussed.

\begin{figure}
\vspace{2mm}
 \vbox to 4.0cm {\vss\hbox to 6.5cm
 {\hss\
   {\includegraphics{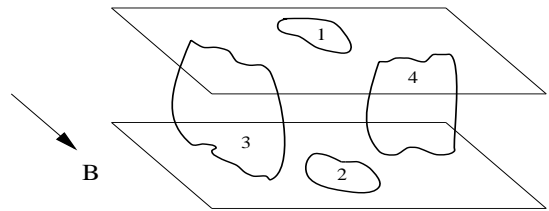}
   }
  \hss}
 }
\caption{Sketch of double quantum wells with in-plane magnetic field.
\label{fig1}}
\end{figure}

Consider an electron travels through
the system of coupled double quantum wells with in-plane magnetic field.
The magnetic field is only relevant if the electron path contains
closed loops 
through inter-layer
hopping (Loops $3$ and $4$ in Fig.1.). 
However, the angles between these loops and the in-plane magnetic
field are random, so the relevant magnetic field components along the direction
perpendicular to these paths are {\it random} and its sign can be
either positive or negative. In another words, this system 
topologically maps into a 2D random field problem. Those paths staying within 
each layer (loops 1 and 2 in Fig.1) or not containing closed loops
correspond to regions with zero field in the random field problem.

According to the effective field theory put forward by Zhang and 
Arovas\cite{za} for the 2D random field problem, a long range logarithmic
interaction arises between the topological density of the nonlinear $\sigma$
model, or more physically the edge excitations of the fluctuating magnetic
domains in the random magnetic field case; and in the coupled double
quantum well system, they should be the hopping paths with different 
orientations regarding to the in-plane magnetic field. They demonstrated that
the 2D random field problem at zero temperature could map into a 2D XY model
at finite temperature\cite{za} in which the longitudinal conductance 
corresponds
to the inverse of the temperature. As opposed to the 2D XY 
model\cite{thouless}, there should
be a disorder driven (where strong disorder corresponds to small conductance,
hence high temperature) Kosterlitz-Thouless transition in 
2D random field problem, or in our case, 
the system of coupled double quantum wells with
in-plane magnetic field at zero temperature. 
Direct consequence of such 
transition is a line of fixed points with continuously varying conductance
in the metallic regime\cite{za}. Using a transfer matrix technique and 
finite-size scaling analysis, we directly calculate the localization length
in the system to demonstrate the existence of the line of fixed points and
critical behavior consistent with KT transition.

In the following, we briefly outline our model and
the technique we used to calculate
the finite-size localization length($\lambda_M$) and to extrapolate the 
thermodynamic localization length($\xi$). 
We model our bi-layer system as two square lattice with interlayer
hopping $t_p$. The square lattice is
a very long strip geometry  with a finite
width ($M$) square lattice with nearest neighbor hopping. Periodic
boundary condition in the width direction is assumed.
The disorder potential is modeled by the on-site
white-noise potential $V_{iml}$ ($i$ denotes the column index, $m$
denotes the chain index, $l$ is the layer index) ranging from $-W/2$ to $W/2$.
The effect of the in-plane magnetic field appears in the complex phase of 
the inter-layer hopping term. 
The strength of the magnetic field is characterized by the flux per
plaquette ($\phi$) in unit of magnetic flux quanta ($\phi_o=hc/e$).
The Hamiltonian of this system can be written as:
\begin{eqnarray} {\cal H} &= &\sum_{l=1,2}\sum_i\sum_{m=1}^{M}
V_{iml}|iml><iml|  \\
& &-t\sum_{l=1,2}\sum_{<im;jn>}\left[
|iml><jnl|+|jnl><iml|\right]\nonumber \\
& &-\sum_{im}\left[ t_p(i)|im1><im2|+t_p^{\dagger}(i)|im2><im1|\right],
\nonumber
\end{eqnarray} 
where $<im;jn>$ indicates nearest neighbors on the
lattice. The amplitude of the in-layer
hopping term($t$) is chosen as the unit of
the energy.  For a specific energy 
$E$, a transfer matrix $T_{i}$ can
be easily set up mapping the wavefunction amplitudes at column
${i-1}$ and $i$ to those 

\begin{figure}
\vspace{2mm}
 \vbox to 5.5cm {\vss\hbox to 6.5cm
 {\hss\
   {\includegraphics{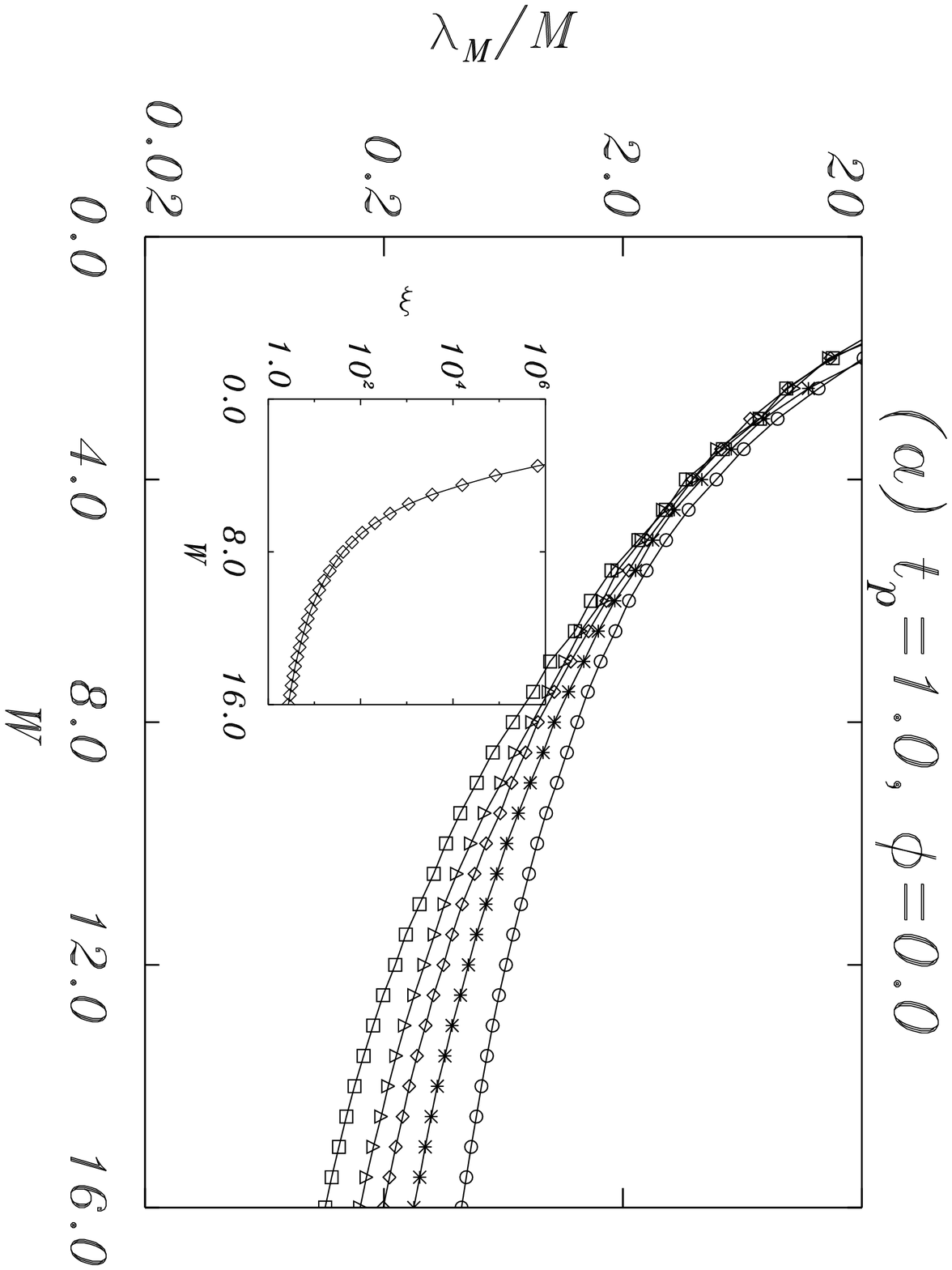}
   }
  \hss}
 }
\vspace{2mm}
 \vbox to 5.5cm {\vss\hbox to 6.5cm
 {\hss\
   {\includegraphics{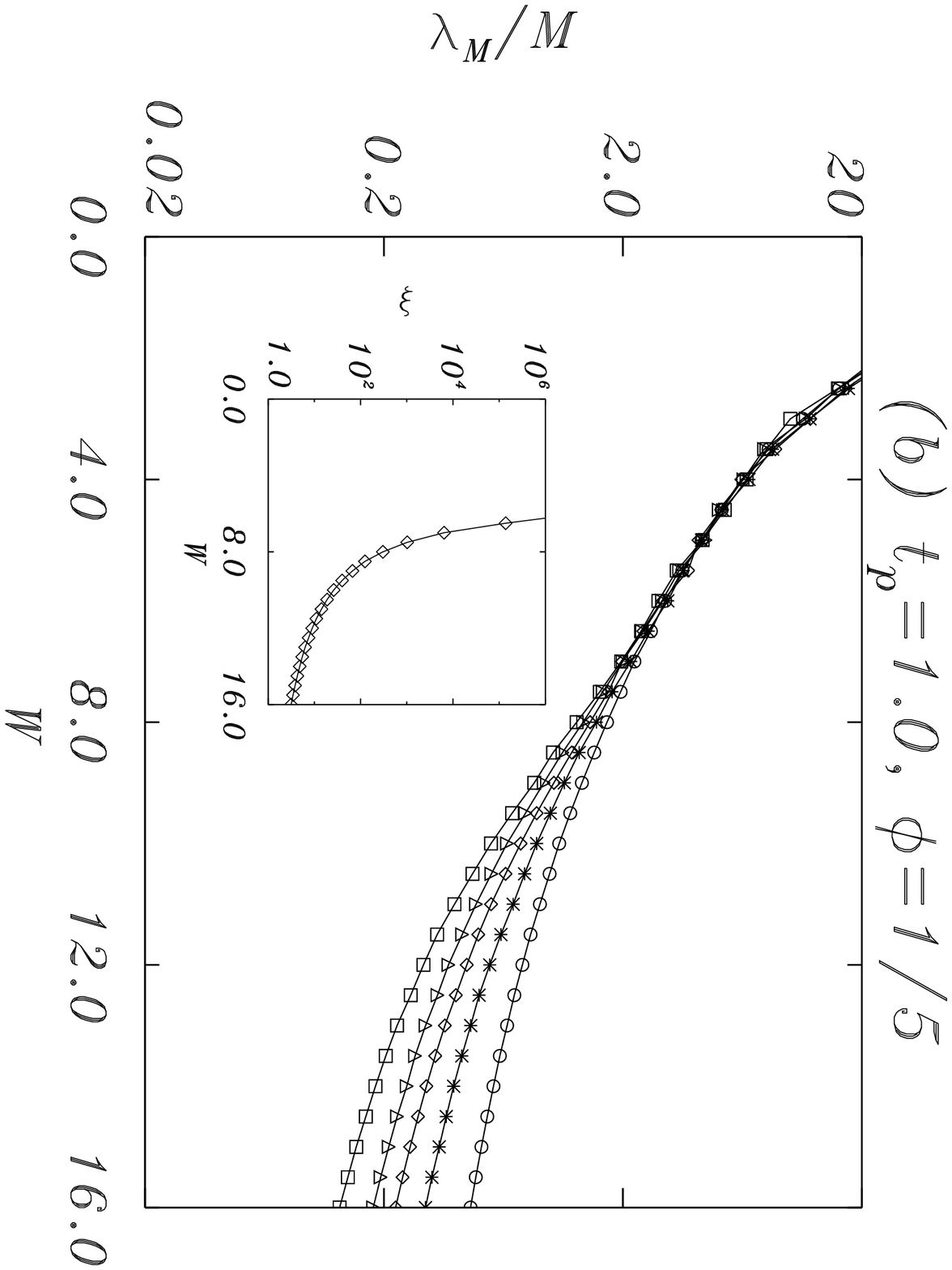}
   }
  \hss}
 }
 \vbox to 5.5cm {\vss\hbox to 6.5cm
 {\hss\
   {\includegraphics{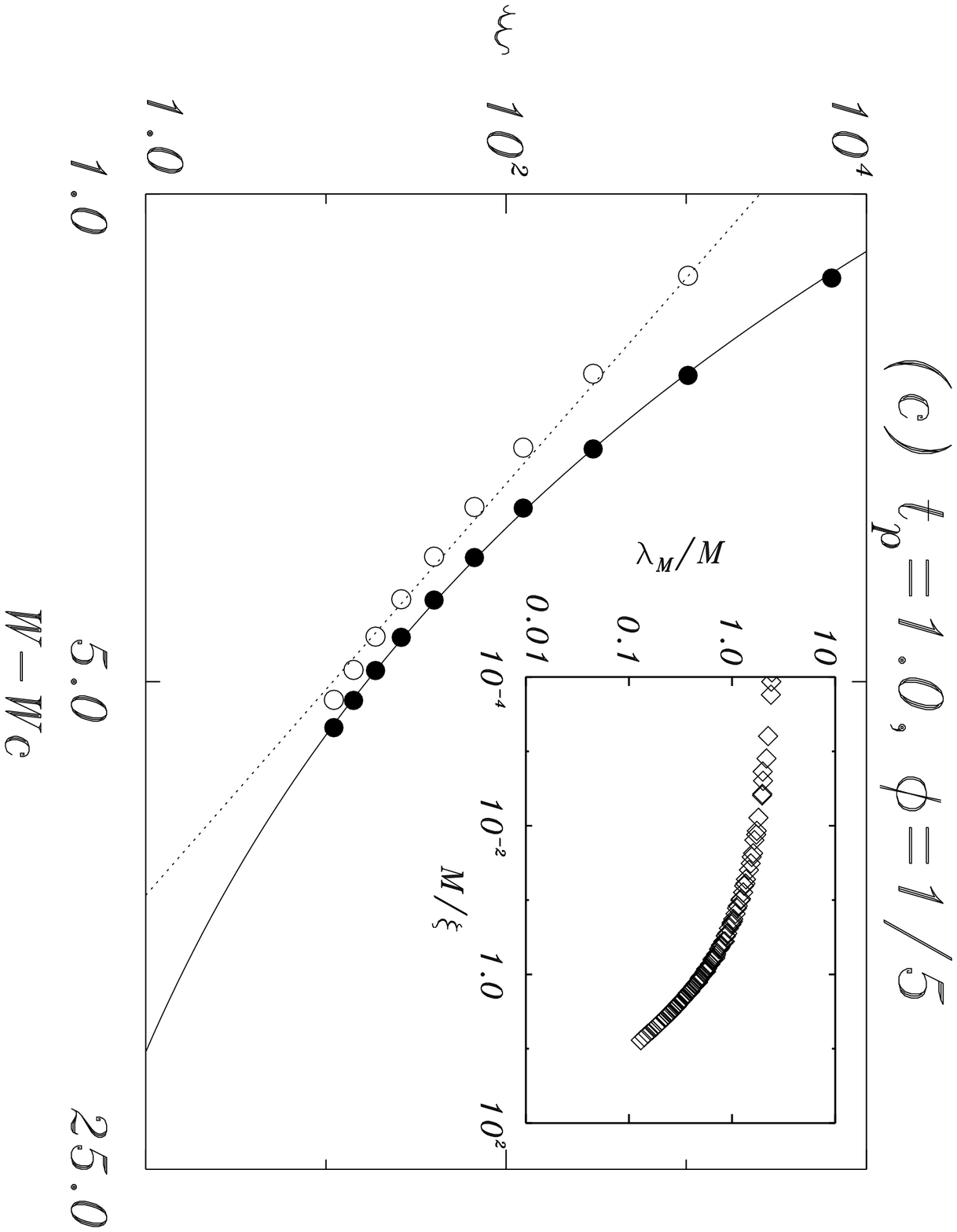}
   }
  \hss}
 }
\caption{
(a) $\lambda_{M}/M$ as function of disorder
strength $W$ without magnetic field for
$M=4$ (circles), $M=8$ (stars), $M=12$ (diamonds),
$M=16$ (triangles) and $M=24$ (squares).
Inset: Thermodynamic localization length $\xi$ as function of
disorder $W$.
(b) $\lambda_{M}/M$ as function of disorder
strength $W$ with magnetic field $\phi =1/5$ for
$M=4$ (circles), $M=8$ (stars), $M=12$ (diamonds),
$M=16$ (triangles) and $M=24$ (squares).
Inset: Thermodynamic localization length $\xi$ as function of
disorder $W$.
(c) Fit the thermodynamic localization length $\xi$ 
with the KT transition and power-law expressions. Inset: scaling function
using the data from Fig.2(b).
\label{fig2}}
\end{figure} 

\noindent
at column $i+1$.
Using a standard iteration
algorithm \cite{ldz:bloc}, 
we can calculate the Lyapunov exponents
for the transfer matrix $T_{i}$.
The localization length $\lambda_M(E)$ for energy
$E$ at finite width $M$ is then given by the inverse of the smallest
Lyapunov exponent. In our numerical calculation, we choose the sample
length to be over $10^5$ so that the self-averaging effect
automatically takes care of the ensemble statistical fluctuations.
We use the standard one-parameter finite-size scaling
analysis\cite{kramer} to obtain the thermodynamic localization length
$\xi$. The scaling calculations are carried out with varying disorder
$W$ (or in-plane magnetic field $\phi$, or inter-layer hopping amplitude
$t_p$) at fixed energy. 

We now discuss our numerical results in various situations and their
implications. Presented in Fig.2(a) is the ratio of
finite localization length $\lambda _M$ with system width $M$
versus disorder $W$ without in-plane magnetic
field. Energy is set at the band center with $E=0$.
Different curves are for different system widths(M=4-24).
Within numerical fluctuations, the curve for smaller $M$ is
above that of the larger $M$ for the entire range of $W$.
This indicates that there is no metal-insulator transition
for finite disorder $W$, consistent with one's expectation
for the double-layer system with zero magnetic field.
In the inset, we show the
thermodynamic 
localization length $\xi$ as function of disorder strength $W$.
Although $\xi$ rises fast for small $W$, numerical fitting
still shows that the critical value $W_{c} \simeq 0$.

Fig.2(b) shows the main results of our paper.
Here we plot the same curves as in Fig.2(a) but with finite
in-plane field $\phi =1/5$. The striking difference with
Fig.2(a) is that all curves merge together for $W < W_c\simeq 5.5$.
As is well known from the finite-size scaling studies of
phase transition, all curves for different sizes should
cross at single point (critical point) for a conventional
continuous 
transition\cite{barber}. Two examples for such transition in the
localization problem are the three-dimensional Anderson
model\cite{kramer} and two-dimensional case with spin-orbital
interaction\cite{spin}. However, Fig.2(b) is quite different
from 
the conventional transition in the sense that there is no
single crossing point, but all curves merge together for
$W<W_{c}$. This shows that all the points for $W<W_{c}$ are
critical points\cite{barber}, namely, we have a line of
critical points which is consistent with our argument that the system 
undergoes a disorder driven Kosterlitz-Thouless transition.
In Fig.2(c), we fit the data for $\xi$ with
$\xi \propto exp(\alpha / \sqrt{W-W_{c}})$ (full circles and solid line), 
typical for the KT transition, and
$\xi \propto (W-W_{c})^{\nu}$ (open circles and dotted line) for $W>W_{c}$.
For the KT fitting, $W_{c}= 5.7$ and $\alpha =13.8  \pm 0.1$
and for the power-law fitting, $W_{c}= 6.2$ and $\nu= 3.4\pm 0.3$.
Results clearly show that KT dependence has a
better fitting and supports the notion that the transition
we found in Fig.2(b) belongs to KT transition.

In Fig.3 we show further numerical support for KT transition
by ploting $\lambda_{M}/M$ versus interlayer hopping $t_p$
for fixed disorder. Fig.3(a) is for zero magnetic field and we see
that curve with smaller $M$ has larger value of $\lambda_{M}/M$
for all $t_p$,
consistent with the picture that all states are localized.
For finite field with $\phi =1/5$ in Fig.3(b), we see all curves
merge together in the range $t_{p}\sim 0.1-3.0$, indicating that
we have continuous critical points in this range.

\begin{figure}
\vspace{2mm}
 \vbox to 5.5cm {\vss\hbox to 6.5cm
 {\hss\
   {\includegraphics{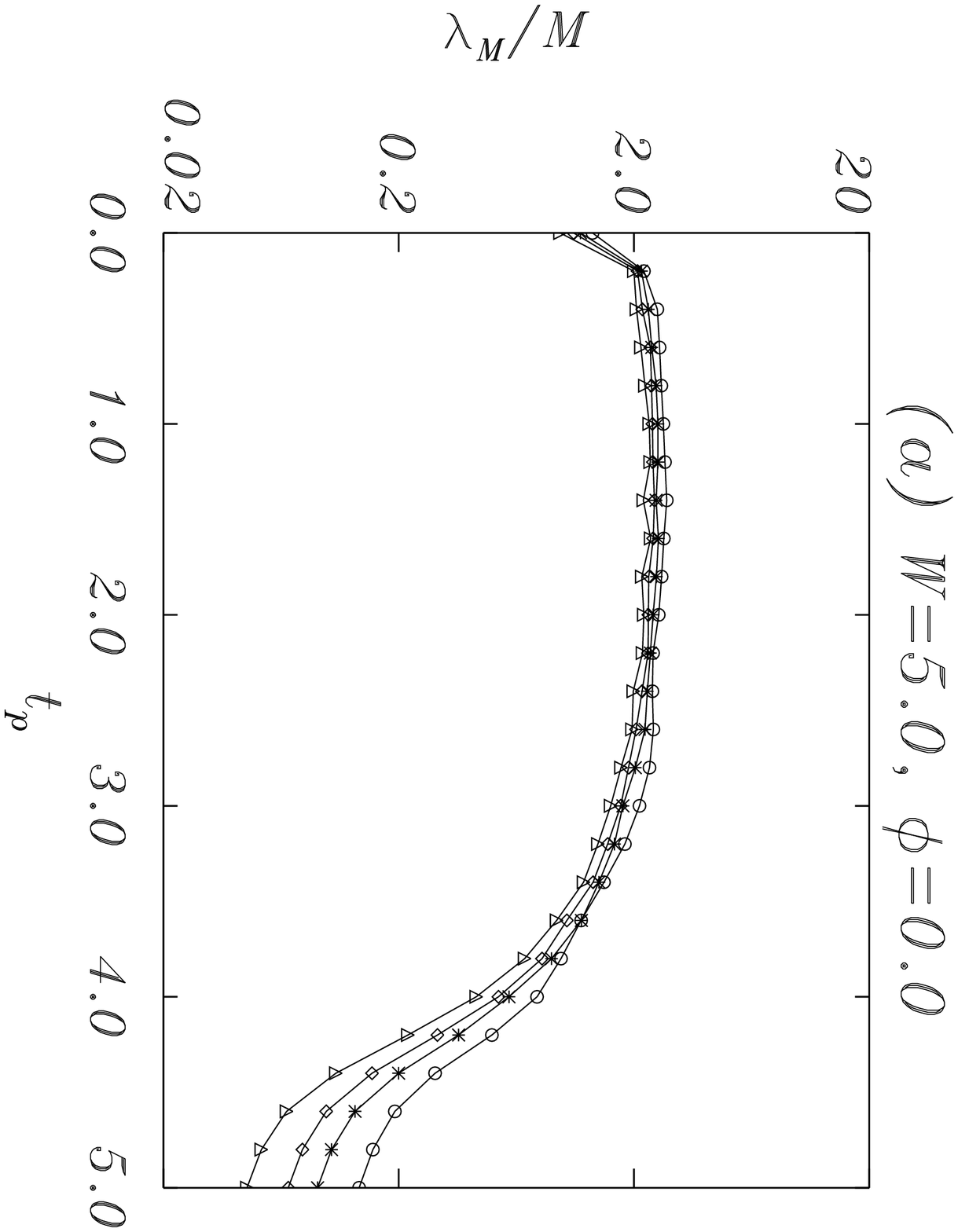}
   }
  \hss}
 }
\vspace{2mm}
 \vbox to 5.5cm {\vss\hbox to 6.5cm
 {\hss\
   {\includegraphics{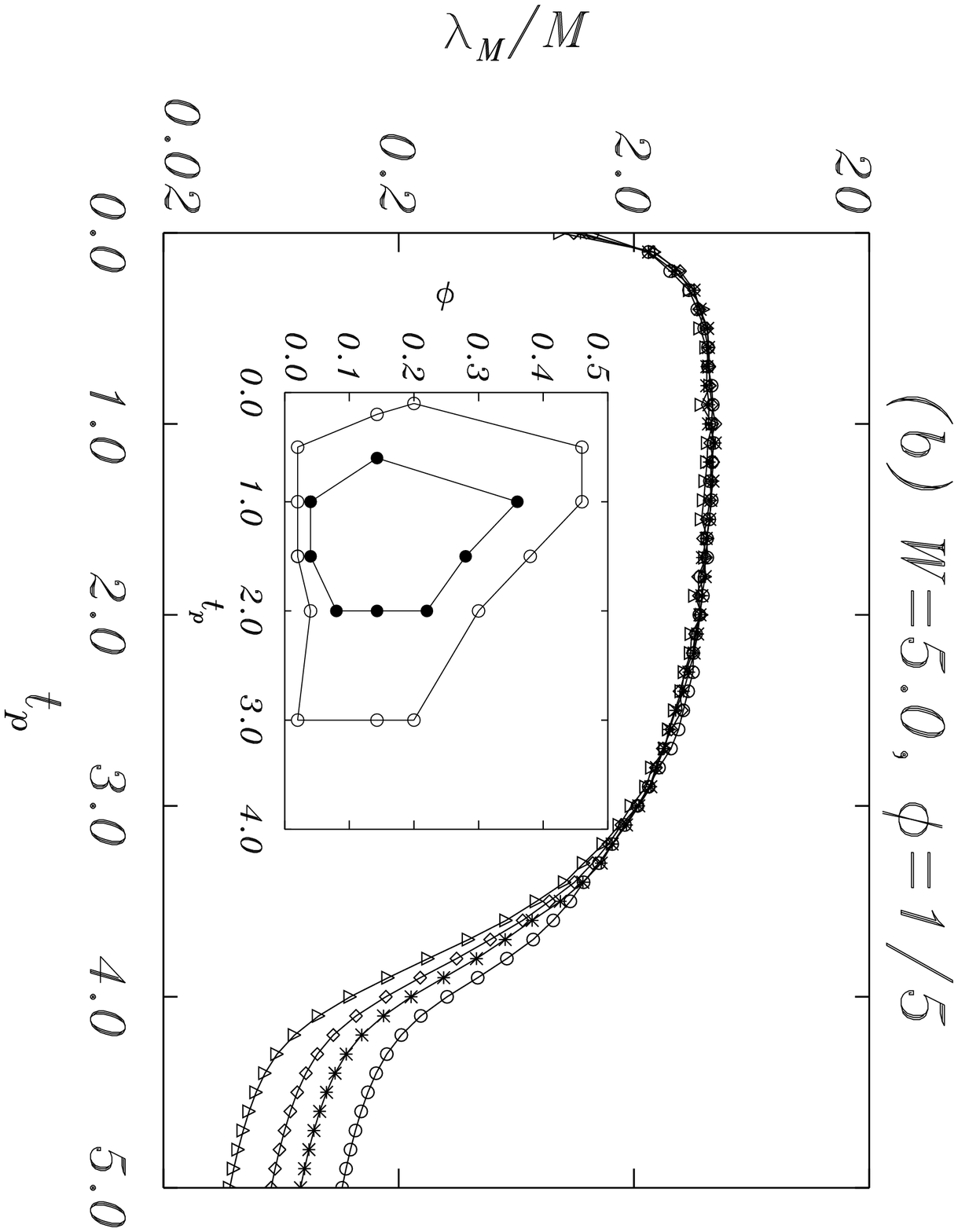}
   }
  \hss}
 }
\caption{
(a) $\lambda_{M}/M$ as function of inter-layer
hopping amplitude $t_p$ without magnetic field for
$M=8$ (circles), $M=12$ (stars), $M=16$ (diamonds),
$M=24$ (triangles).
(b) $\lambda_{M}/M$ as function of inter-layer
hopping amplitude $t_p$
with magnetic field $\phi =1/5$ for
$M=8$ (circles), $M=12$ (stars), $M=16$ (diamonds),
$M=24$ (triangles).
Inset: Phase diagram on the $\phi - t_p$ plane. The filled circles are
for W=6, the opened ones are for W=5. The states within each closed loop 
are extended states.
\label{fig3}}
\end{figure}

In a realistic experimental set-up, the sample disorder is
fixed and the in-plane magnetic field can be easily changed.
In the following we study the phase diagram for fixed disorder
by varying magnetic field and inter-layer hopping which depends
on the distance between the two layers.
For a fixed moderate coupling strength between two layers, in-plane magnetic
has no effect on those electron paths within each layer, these states tend to
stay localized. However, for those paths going through inter-layer hopping,
the time-reversal symmetry is broken. So as the magnitude of magnetic field
increases from zero, the system undergoes a metal-insulator transition at
a critical magnetic field. In the presence of
 extremely strong magnetic field, the magnetic
length is comparable to the inter-layer separation, the system will become
localized again with all electrons quenched within the magnetic length.
The story is similar
in a fixed moderate magnetic field. When the inter-layer coupling is very
small, the number of paths going through inter-layer hopping is very limited.
So the system behaves as two separate layers and all states are localized.
When the coupling strength is very large, the energy gap between the
symmetric and antisymmetric states is very large. Thus, the hopping
between the symmetric and antisymmetric states is very weak, and
again all states are localized. Extended states only emerge in the
intermediate coupling regime. 
We have carried out numerical calculations 
for many points on the $\phi-t_p$ plane.
Presented in the insert of Fig.3(b) is the resulting phase boundary separating
the metallic and insulator regime. 
The phase diagram is determined by the points
when those lines of $\lambda_{M}/M$ for different $M$ merge together.
Indeed the metallic regime is a window 
on the $\phi-t_p$ plane and it shrinks when the system gets more disordered.

\begin{figure}
\vspace{2mm}
 \vbox to 5.5cm {\vss\hbox to 6.5cm
 {\hss\
   {\includegraphics{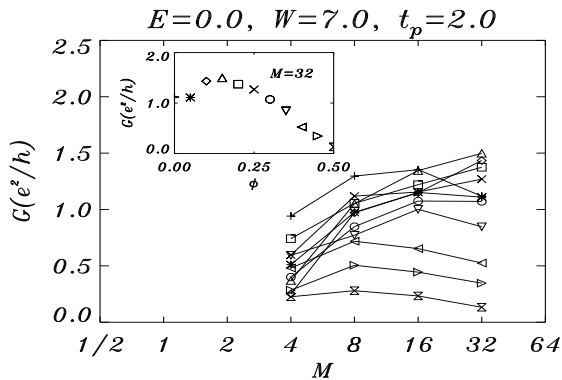}
   }
  \hss}
 }
\caption{
Average conductance (over 100 samples) $G$ as function of sample size.
Different symbol represents different in-plane magnetic field as shown
in the inset (field dependent conductance for $M=32$).
\label{fig4}}
\end{figure}

The above phase diagram can be tested by measuring the temperature
dependence of the conductance in coupled double quantum wells with
in-plane magnetic field. We mimic this effect by calculating the sample
size dependence of such conductance, since finite temperature provides
an effective cut-off length (the inelastic scattering length) in a real 
system. The conductance is calculated using the Landauer-B\"{u}ttiker formula
and recursive 
Green's function technique.  As presented in Fig.\ref{fig4},
one can see that the conductance in the metallic regime increases
with the sample size(mimic decreasing temperature), 
while that for insulator regime decreases. 
The conductance should reach maximum in the extended regime while varying 
the in-plane field at fixed sample size(fixed temperature) as presented in
the inset.  
To test the effect of the coupling strength, one could do similar experiment
on different samples with various barrier height and thickness in the
double quantum wells.

In summary, we have demonstrated that the system of  double quantum wells
with in-plane magnetic field maps into the 2D random field problem and hence
a 2D XY model, which undergoes a disorder driven KT type 
metal-insulator transition depending the inter-layer coupling and the magnetic
field. In fact, the metallic regime is a window on the field--coupling plane
which shrinks with increasing disorder. We propose temperature dependence 
transport measurement to test this phase diagram.

We thank Song He and Michael Ma for many helpful discussions.
X.C. Xie is supported by NSF-EPSCoR.

\vspace{-0.5cm}

\end{document}